\documentclass[10pt,superscriptaddress,reprint,amsmath,amssymb,aps,prl,showpacs]{revtex4-2}
\usepackage{mathrsfs}
\usepackage{graphicx}
\usepackage[caption=false,farskip=0pt,labelfont={bf}]{subfig}
\usepackage{dcolumn}
\usepackage{bm}
\usepackage{amssymb}
\usepackage{amsmath}
\usepackage{paralist}
\usepackage{amsmath,amssymb,mathrsfs}

\usepackage[colorlinks,linkcolor=blue,anchorcolor=blue,citecolor=green]{hyperref}
\usepackage{cleveref}
\usepackage{url}
\usepackage{color}
\usepackage{subfig}

\makeatletter

\renewcommand{\maketag@@@}[1]{\hbox{\m@th\normalsize\normalfont#1}}%

\makeatother

\begin{document}

\title{Dynamical Correlation of the Post-quench Non-thermal Equilibrium State}

\author{Yang-Yang Chen}
\affiliation{Institute of Modern Physics, Northwest University, Xi'an 710069, China}
\affiliation{Shaanxi Key Laboratory for Theoretical Physics Frontiers,  Xi'an 710069, China}

\author{Song Cheng}
\email[]{song919@hku.hk}
\affiliation{Department of Physics, The University of Hong Kong, Hong Kong, China}



\begin{abstract}

After a quantum quench, the integrable system is expected to relax to a non-thermal equilibrium state (NTES) whose local properties are believed to be governed by a generalized Gibbs ensemble (GGE). 
Combining quench action and the form factor approach, we compute the field-field correlation in the NTES produced by an interaction quench of the Lieb-Liniger model. 
The spectral distribution is shown to be qualitatively different from that of a thermal equilibrium state (TES): 
a new dispersion branch appears whose microscopic mechanism can be traced to the algebraic decaying tail for the root density distribution function, and indicates the existence of a broader family of NTES featuring similar spectral property.

\end{abstract}

\maketitle

\textit{Introduction.}--- The unitary time evolution of isolated quantum systems has been a fruitful research field in decades, due to the convenient engineering of ultra-cold atom systems \cite{Bloch2008,Nori2014,Guan2022} and its intimate connection with statistical mechanics \cite{Srednicki1994, Rigol2012,Rigol2016,Eisert2015,Eisert2016}.
A paradigmatic setting is the quantum quench: the system prepared in an initial state $|\Psi_I\rangle$ evolves under a Hamiltonian $\hat{H}$, for which $|\Psi_I\rangle$ is not an eigenstate.
In the thermodynamic limit, the system is expected to relax to a stationary state, whose nature according to the choice of $\hat{H}$ and the initial state may differ a lot \cite{Mossel2012,Andrei2012,Kormos2013,Fagotti2013a,Fagotti2013b,Fagotti2014,Piroli2016a,Piroli2016b,Piroli2017a,Piroli2017b,Piroli2019,Alba2016,Alba2017a,Alba2017b,Alba2017c,Alba2018,Bertini2014,Bertini2016,Schuricht2023,Bertini2017,Bertini2022,Bertini2024,Delfino2023,Horvath2024,Rylands2023,Pietraszewicz2019,Cazalilla2009,Schuricht2016,Meden2012,Meden2013,Meden2014a,Meden2014b,Meden2017,Zill2015,Zill2018,Essler2015,Caux2013,Brockmann2014,Caux2016,DeLuca2015,Bucciantini2016,Nardis2014a,Nardis2014b,Sotiriadis2014,Pozsgay2014a,Caux2014,Pozsgay2014b,Prosen2015,Rigol2016,Zhang2024}. 
Specially, for an integrable Hamiltonian possessing infinite conserved charges, present in the late-time limit is a non-thermal equilibrium state (NTES) typically representing a generalized Gibbs ensemble (GGE) \cite{Sotiriadis2014,Pozsgay2014a,Caux2014,Pozsgay2014b,Prosen2015,Rigol2016}. 
In this sense, a quantum integrable system is usually said not to be thermalized.

The physics underlying quantum quench has been elucidated in several key aspects, including the explicit GGE construction \cite{Fagotti2013a,Fagotti2013b,Piroli2017a,Essler2015}, dynamical relaxation \cite{Mossel2012,Andrei2012,Fagotti2014,Alba2017a,Alba2017b,Alba2017c,Alba2018,Bertini2014,Zill2015,Zill2018,Nardis2014b}, and equal-time correlation functions in the NTES \cite{Kormos2013,Fagotti2013b,Piroli2016a,Piroli2016b,Piroli2017a,Rylands2023,Bertini2016,Bertini2022,Zill2015}, etc.
By contrast, dynamical correlation functions (DCFs) in the NTES are much less understood, especially for genuinely integrable models that cannot be mapped onto free particles.
The rapid progress of experimental spectroscopy \cite{Fabbri2015,Li2024,Senaratne2022,Hulet2023,Yu2025} now places this gap in sharp relief: can a simple spectroscopic measurement discriminate between an NTES and a TES produced by a quantum quench? To answer this fundamental question, the detailed distribution of spectral weights is compulsory.

At first glance, these DCFs can be obtained from the GGE partition function together with {\em form factors} of the correlators. 
However, explicit GGE construction has been shown to be quite challenging \cite{Piroli2017a,Sotiriadis2014,Pozsgay2014a,Caux2014,Pozsgay2014b,Prosen2015,Zhang2024}, such as in the presence of string bound states in the spectrum of quenching Hamiltonian. 
The other obstacle lies in implementing the form factor approach to DCFs in a highly excited state necessitating a tremendous number of intermediate states, which is hardly accessible through ABACUS \cite{Caux2006,Panfil,Caux2007,Caux2009,Klerk2023}.
To overwhelm the above difficulty, we combine our recently developed algorithm based on form factors \cite{Cheng2025a,Cheng2025b,Li2023} with the quench action approach \cite{Caux2013,Caux2016,Nardis2014a}.
As a nontrivial example, in this letter we study the field-field correlation in the NTES produced by the interaction quench of the Lieb-Liniger model.
We find that the spectral distribution in this NTES is qualitatively distinct from a TES of the same energetic and interacting parameters and uncover the mechanism of this phenomenon in a microscopic perspective, indicating a family of NTESs possessing similar spectral property for other integrable models and quench protocols \cite{Rylands2023}.

The initial state is chosen to be the non-interacting ground state, and the Hamilton of Lieb-Liniger model is given by
\begin{equation}
\label{ham}
H = -  \sum_{i=1}^{N} \frac{\partial^2}{\partial x_i^2} + 2c \sum_{i>j}^{N} \delta \left( x_i - x_j \right),
\end{equation}
where we denote by $N$ and $c$ the particle number and the interaction strength,  respectively \cite{Lieb,Jiang,QISM}. 
Under periodic boundary condition, the system can be solved exactly by the Bethe ansatz, yielding the following Bethe equations
\begin{equation}
\label{LBAE}
\lambda_j + \frac{2}{L} \sum_{k=1}^{N} \arctan \left( \frac{\lambda_j - \lambda_k}{c} \right) = \frac{2\pi}{L} I_j
\quad j = 1, \dots, N
\end{equation}
where $\lambda$ is called rapidity and $\{I_j\}_N$ the quantum number (QN) is a set of distinct integers (half-odd integers) in the case of odd (even) $N$.
It can be proved that each solution to Eq.~\ref{LBAE} consists of $N$ real distinct roots, denoted by $\{\lambda\}_N$. 
The total momentum and energy of the system can be written in terms of these rapidities
$P_{\{\lambda\}_N}=\sum_{j=1}^{N} \lambda_j$, and $E_{\{\lambda\}_N}=\sum_{j=1}^{N} \lambda_j^2.$
Worth mentioning is the one-to-one correspondence between $\{\lambda_j\}_N$ and $\{I_j\}_N$.

{\em Methods.---}
The {\em form factor} of an operator $\hat{O}$, denoted by $\mathcal{F}(|s\rangle,|r\rangle)=\langle s|\hat{O}|r\rangle$, is defined as the matrix element between two eigenstates of the Hamiltonian \cite{QISM}. Consequently the spectral representation for the time-dependent correlation function in the eigenstate $|s\rangle$ is formulated by
\begin{equation}\label{spectral_representation}
\langle s| \hat{O}^\dagger(x,t) \hat{O}(0,0) | s \rangle = \sum \limits_{|r\rangle \in \mathfrak{E}} e^{\mathrm{i} \varphi} \frac{|\mathcal{F}(|s\rangle,|r\rangle) |^2} { \| s \|^2 \, \| r \|^2},
\end{equation}
where the intermediate states $|r\rangle$ are taken over the entire eigenspace $\mathfrak{E}$, the phase factor $\varphi = E_{r,s}t - P_{r,s}x$, the difference in energy (momentum)  $E_{r,s}=E_r-E_s$  ($P_{r,s}=P_r -P_s$), and $\| \dots \|$ means the norm of a state.
The quantum integrability makes the spectrum accessible, and the QNs  labeling eigenstates here are analogous to the situation for noninteracting spinless fermions, so is the concept of particle-hole (p-h) excitation \cite{QISM,Jiang}.
Obviously, the key to Eq.~\ref{spectral_representation} is how to {\em efficiently} enumerate the elements of $\mathfrak{E}$ together with calculating their form factors.
Moreover, experimentalists usually request an easy access to momentum-wise results, e.g. the dynamical structure factor.

For this purpose, we developed an algorithm based on classifying intermediate states through the generalized p-h excitations over an arbitrarily given eigenstate $|s\rangle$ \cite{Cheng2025a,Cheng2025b,Li2023}. 
Subsequently, $\mathfrak{E}$ is partitioned into a series of equivalent classes through a tag set consisting of four quantum numbers 
$\{P_\mathrm{m},N_\mathrm{p},P_\mathrm{l},N_\mathrm{l}\}$, where $P_\textrm{m}$ ($N_\textrm{p}$) represents the given excited momentum (the number of generalized p-h excitations) and $P_\textrm{l}$ ($N_\textrm{l}$) represents the momentum (the number) of the excitations towards the negative direction. All momenta are measured in units of $2\pi/L$.
Note that $P_\mathrm{m}, N_\mathrm{p}, P_\mathrm{l} \in \mathbb{N}$ are mutually independent, while $N_\mathrm{l} \in [ 1, \mathrm{min}(N_\mathrm{p}-1,P_\mathrm{l})] \bigcap  \mathbb{N}$ with $\mathbb{N}$ being the set of natural numbers.
The practical implementation starts with a given tag set, then produces and incorporates the states belonging to this equivalent class; generates a new tag set by gradually increasing quantum numbers shown in the precedent tag set and repeats precedent treatment until the criterion of truncation is achieved either momentum- or energy-wise; see the Supplementary Material.

\begin{figure}
    \centering
    \includegraphics[width=0.9\linewidth]{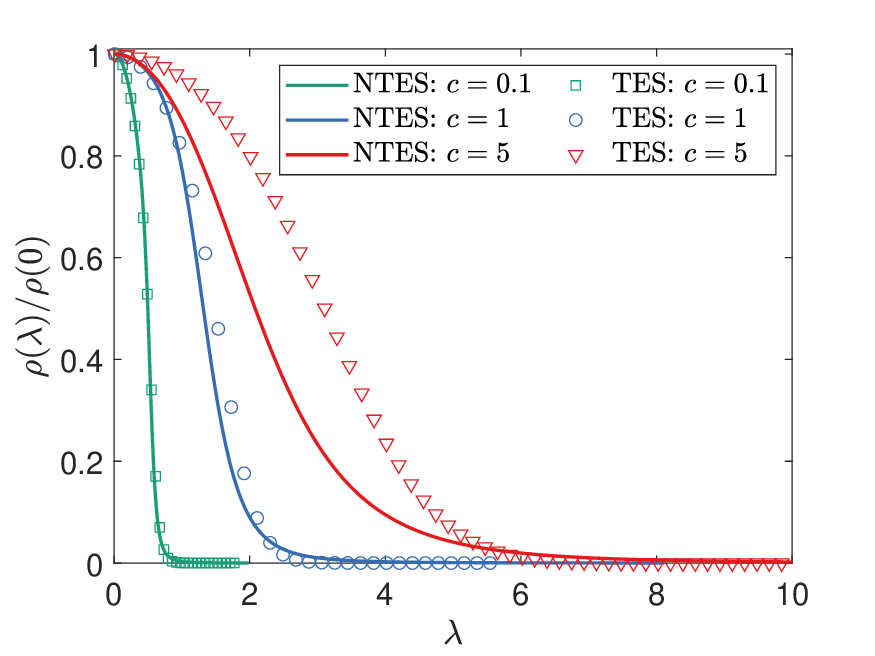}
    \caption{The rescaled root density distribution $\rho(\lambda)/\rho(0)$ are shown for TES and NTES of different values of $c$. The solid line and symbol reresent the results of NTES and TES, respectively.}
    \label{fig_compare_rho}
\end{figure}

In the thermodynamic limit (TL), an eigenstate can be expressed macroscopically in the rapidity space, i.e. $\rho(\lambda_j)= \lim\limits_\textrm{TL} \frac{1}{L(\lambda_{j+1} - \lambda_j)}$.
Specific to an equilibrium state, it satisfies 
\begin{equation}
    \rho(\lambda) = \frac{1}{2\pi} \vartheta(\lambda) \left[1 +  K \ast \rho (\lambda)\right],
\end{equation}
with the convolution operator denoted by $\ast$ and the kernel function by $K(x,y)=\frac{2c}{c^2+(x-y)^2}$. Here $\vartheta(\lambda)=\frac{1}{1+e^{\epsilon(\lambda)}}$ is the filling function defined through the thermodynamic Bethe ansatz equation (TBA) that the system obeys \cite{YangYangThermo}. 
For a TES, it is given by
\begin{equation}
    \epsilon_\textrm{TES} (\lambda) =  \frac{\lambda^2-h}{T} - \frac{1}{2\pi} K \ast \ln\left[ 1 + e^{-\epsilon_\textrm{TES}}\right] (\lambda),
\end{equation}
with chemical potential  $h$  \cite{YangYangThermo,Jiang,QISM}. For the NTES under study, a generalized TBA is written as
\begin{align}
    \epsilon_\textrm{NTES} (\lambda) = & \ln \left[ \frac{\lambda^2}{c^2} \left( \frac{\lambda^2}{c^2} + \frac{1}{4} \right) \right] - h' 
    \notag  \\
    &- \frac{1}{2\pi} K \ast \ln\left[ 1 + e^{-\epsilon_\textrm{NTES}}\right] (\lambda),
\end{align}
with Lagrange multiplier $h'$, which was first derived in \cite{Nardis2014a} through the quench action approach. 
The total number of particles or the density is imposed through 
$\int_{-\infty}^\infty \textrm{d} \lambda \, \rho(\lambda) = N/L$.
%
We compare $\rho_\textrm{TES}(\lambda)$ and $\rho_\textrm{NTES}(\lambda)$ of different interaction strengths but with fixed density $N/L=1$ and the same energy density $E/L=\int_{-\infty}^\infty \textrm{d}\lambda\,  \lambda^2 \rho(\lambda)$
in Fig.~\ref{fig_compare_rho}, where they are represented by symbols and solid lines, respectively.

The calculation for $\mathcal{F}(|s\rangle,|r\rangle)$ will be carried out for a finite-size system, 
and thus we need the eigenstate, alias the discrete QN in conformity with $\rho_\textrm{NTES}(\lambda)$.
Through a sampling process, we acquire a series of such eigenstates of quasi-degeneracy, which can be roughly thought of as copies.
In the following, we consider the correlation of the NTES over all these copies.

\begin{figure}[htbp]
\centering
\includegraphics[width=1.05\linewidth]{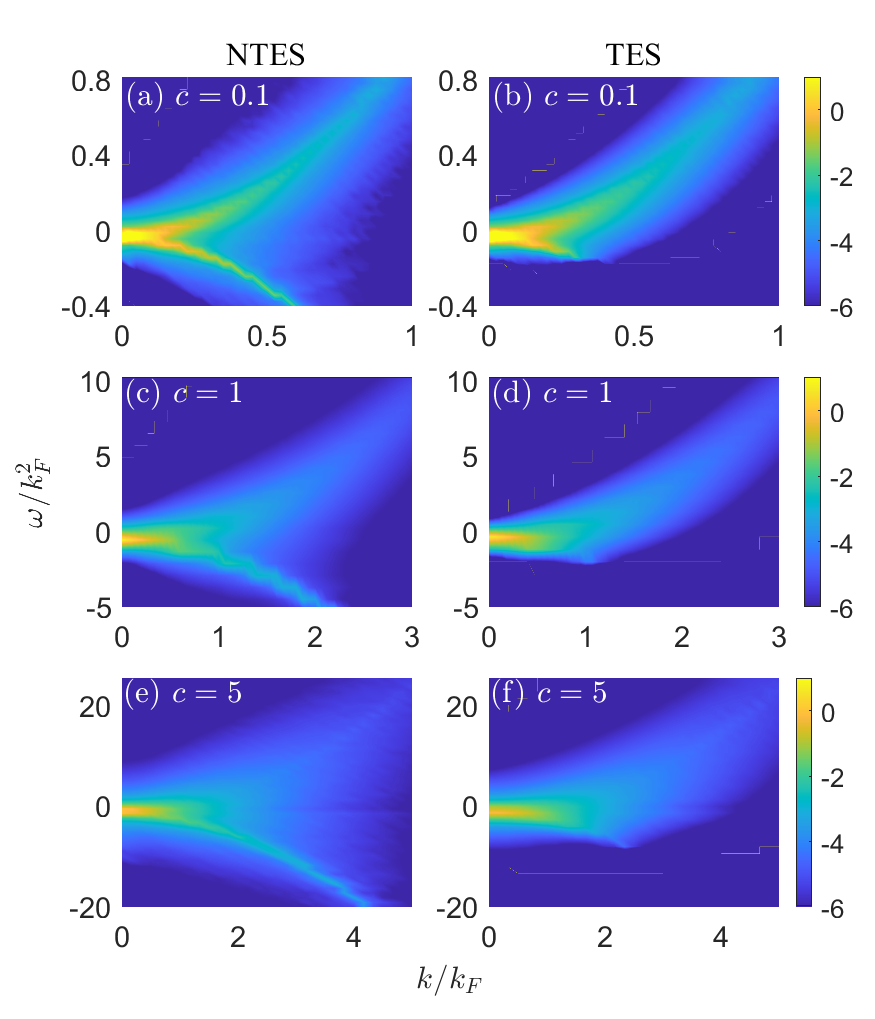} 
\caption{The logarithm of 1BDCF in NTES and TES in the momentum-energy plane.
The momentum and energy are measured in units of Fermi momentum $k_F = \pi N/L$ and Fermi energy, respectively. 
From top to bottom, the interaction varies from weak $c=0.1$, medium $c=1$ to strong $c=5$; the corresponding system size is $N=L=80$, $30$ and $12$; the temperature of TES takes $T=0.13$, $1.15$ and $6.36$, respectively.
The sum rules, c.f. Eq.~\ref{sumrule}, of (a) to (f) are 99.99\%, 99.99\%, 99.99\%, 99.97\%, 99.72\%, and 99.99\%, respectively.}
\label{fig_contour}
\end{figure}

\textit{The 1BDCF in the $k$-$\omega$ plane.}------
We take the field-field correlation as a non-trivial example to tackle the DCF in an NTES. Its Fourier transform is called the 
one-body dynamical correlation function (1BDCF) \cite{Caux2006,Cheng2025a}, defined by
\begin{small}
\begin{equation}
    g_1(k,\omega) = \int_0^L \textrm{d}x \int_{-\infty}^\infty \textrm{d}t \,e^{\textrm{i}(kx-\omega t)} 
    \langle  \Psi^\dagger(x,t)  \Psi(0,0)\rangle,
\end{equation}    
\end{small}%
with the bosonic field operator $\Psi^\dagger(x,t)$. 
The 1BDCF satisfies a sum rule as follow
\begin{equation}\label{sumrule}
    \frac{N}{L} = \sum_k \int_{-\infty}^\infty \frac{\textrm{d}\omega}{2\pi L} \, g_1(k,\omega),
\end{equation}
which is a useful criterion when checking the validity of the results obtained.
We set $N=L$, and thus the closer we get to unity, the better the sum rule.

\begin{figure}[htbp]
\centering
\includegraphics[width=1.0\linewidth]{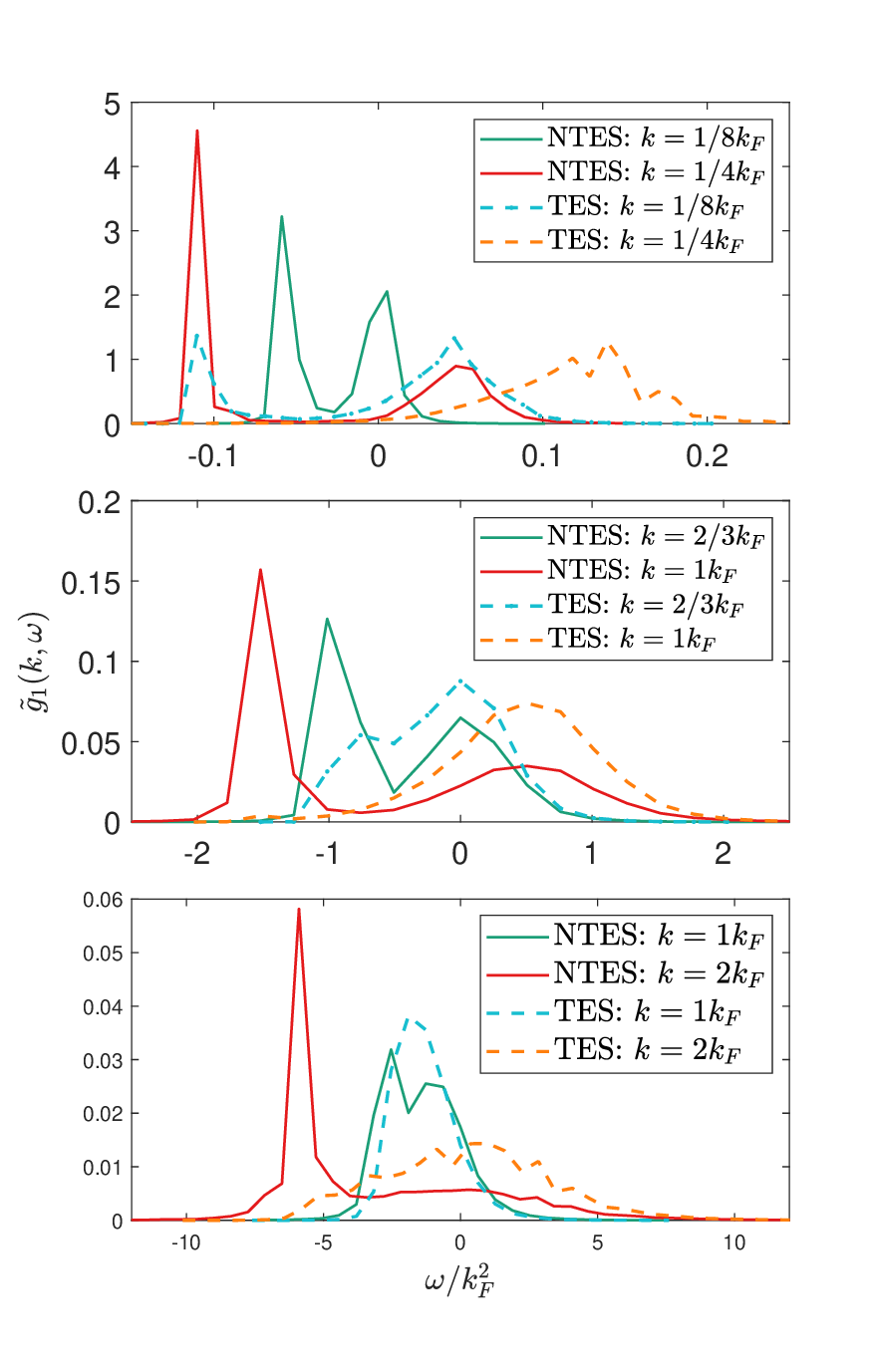}
\caption{Plot of function $\tilde{g}_1(k,\omega)$ with argument $\omega$ for several different momenta. 
The results for TES and NTES are respectively represented by solid and dashed lines.
The interaction strength $c$ takes $0.1$, $1$, and $5$ from top to bottom.}
\label{fig_line}
\end{figure}

Using our recently developed algorithm \cite{Li2023,Cheng2025a,Cheng2025b}, we plot the 1BDCF for NTES suffering from interaction quenches of different strengths in the left panel of Fig.~\ref{fig_contour}. For comparison, the results for TES with the same interaction strength and energy density are shown in the right panel.
It is simple to see the qualitative difference between two panels: the NTES displays a lower dispersion in the negative energy plane that is entirely absent in the TES, cf. Fig.~\ref{fig_line} as well.
To our consternation, although the root density distributions, $\rho_\textrm{NTES}(\lambda)$ and $\rho_\textrm{TES}(\lambda)$, look similar for the weak coupling, cf. Fig.~\ref{fig_compare_rho},
their dynamical correlations are completely dissimilar.
It should be stressed that this dispersion is neither type-I nor type-II dispersion for the ground state of \cite{Lieb}, not even their remnants in low-lying excited states.
What we treat here are highly excited states, and thus there is no longer a well-defined Fermi sea \cite{QISM,Jiang}, leading to the abortion of conventional particle or hole-like dispersion.
To unveil the origin of this new dispersion, let us analyze $\rho_\textrm{NTES}(\lambda)$ and $\rho_\textrm{TES}(\lambda)$. 
It is legitimate to notice their key difference in the tail:
the former decays in a power law given by $\lim\limits_{\lambda \rightarrow \infty} \rho_\textrm{NTES}(\lambda) \sim 1/\lambda^4$, which in Tonks-Girardeau limit becomes $1/\lambda^2$ \cite{Kormos2013,Nardis2014a};
meanwhile, the latter shows an exponential decay.
Consequently, this leads to the following QN configuration: the QNs of TES tend to stay closer than the NTES's.
In common parlance, very few exterior particles, in the QN configuration for an NTES, are located far from the rest and move like \textit{free} particles, which must be reckoned with when sorting the intermediate states bringing in significant spectral weights.

\begin{figure}[hbpt]
    \centering
    \includegraphics[width=1.0\linewidth]{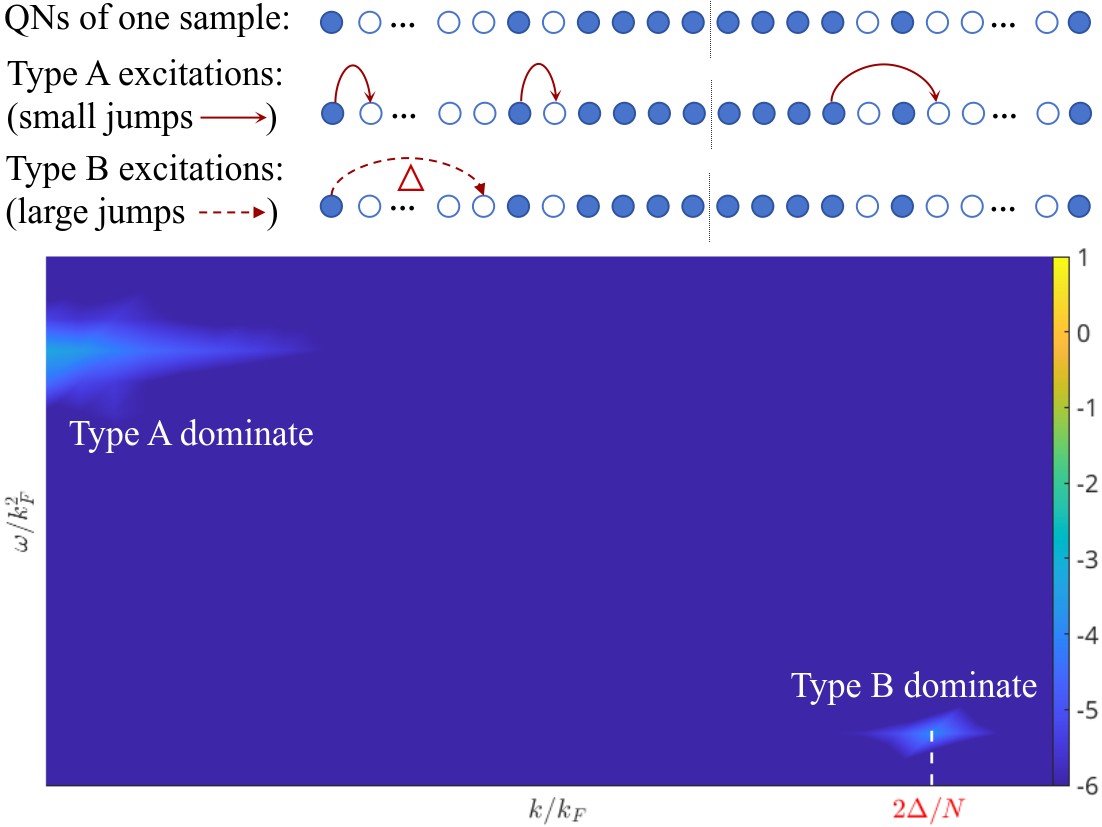}
    \caption{The 1st line shows the QN configuration for a sample of the NTES where the solid and hollow ball represents the particle and hole, respectively; following two lines demonstrate how the intermediate states are produced by a combination of generalized p-h excitations. The bottom plot shows the most spectral distribution resulted from above two categories of intermediate states produced by corresponding types of excitations.}
    \label{fig:example}
\end{figure}

We sketch how the lower dispersion is raised in  a sample of NTES when $c=5$ in Fig.~\ref{fig:example}.
Therein, the 2nd line exhibits the excitation bearing small momenta, which are undertaken by the left-most particle moving towards the center of Fermi sea, together with other right-ward excitations. 
They collectively form the Type A intermediate states.
Of special interest is the excitation shown by the 3rd line, where a single-particle excitation of its own, called Type B, marks the lower dispersion with finite momentum restricted by $k/k_F \lesssim 2\Delta/N$, where $\Delta$ is the distance between the two left-most particles in the QN configuration of the sample state.
The type B excitation launches the foundation of the lower branch of dispersion.
The most shiny spectral contributions from two categories of intermediate states are clearly plotted in the bottom of Fig.~\ref{fig:example}, where the two most bright regions are labeled by the dominant type of excitations. 
The spectral weights in other regions are less significant and the continuous dispersion shown in Fig.~\ref{fig_contour} comes from a superposition of many copies.

\textit{The line shape of 1BDCF.}------
In a TES, the spectral weight is confined mainly to the half plane of positive energy, and the spectral distribution is quantitatively governed by the detailed balancing relation (DBR) \cite{Lifshitz,Cheng2025a}.
Until the system approaches the high temperature limit, the peak located in the positive energy is generally more manifest than in the negative case.
For the GGE, a generalized DBR is expected. 
However, specific to the current quench problem, all attempts to construct a GGE so far have failed due to the obstacle of the power-law tail for the root density distribution \cite{Nardis2014a,Caux2016,Kormos2013}, and so does a generalized DBR.
Therefore, a thorough investigation of the line shape for a DCF is in a urging demand. 
For easy visibility, we introduce a rescaled 1BDCF as follows 
\begin{equation}
    \tilde{g}_1(k,\omega) = \frac{g_1(k,\omega)}{\int_{-\infty}^\infty g_1(k,\omega) \textrm{d}\omega},
\end{equation}
and plot it for several momenta in Fig.~\ref{fig_line}, with parametric settings the same as in Fig.~\ref{fig_contour}.

The lower dispersion for the NTES is prominently evidenced by the peaks located in negative energies, and the peaks in positive energies symbolizing the upper dispersion, although not as obvious as the lower case, are also observed.
The direct and exact calculation of the line shape for $\tilde{g}_1$ signals the qualitative difference in the spectral distribution between NTES and TES: for NTES, the left peak is higher than the right, contrary to the situation of TE; as the momentum grows, the double-peak structure of 1BDCF in TES fades away, while it survives in NTES.

\textit{Discussion \& Conclusion.}------
By combining our recently developed algorithm based on form factors with the quench action approach, we successfully computed the 1BDCF in the $k$-$\omega$ plane for the NTES following an interaction quench in the Lieb-Liniger model. Our results clearly demonstrate a qualitative difference between the NTES and a TES with the same energy density and interaction strength: specifically, the emergence of a lower dispersion branch. We explain the microscopic origin of this feature by analyzing the QN configuration, highlighting the critical role played by a small number of exterior particles governed by the power-law tail of the root density $\rho_{\text{NTES}}(\lambda)$ as $\lambda \to \infty$. This explanation is corroborated by an analysis of the spectral weight contributed by two categories of intermediate states arising from the corresponding excitation types. We argue that these findings suggest the existence of a broader family of NTESs sharing similar spectral properties. For instance, in the BEC-BCS quench of the attractive Gaudin-Yang model \cite{Rylands2023}, where the root density distribution functions also decay via a power law $\rho(k) \sim 1/k^4$ and $\tilde{\sigma}(\lambda) \sim 1/\lambda^4$, it is reasonable to conjecture that a similar lower dispersion branch will appear. 
In the future, advances in the experimental realization of box potentials and spectroscopic techniques will facilitate the direct measurement of dynamical correlation functions in NTES. 
Furthermore, categorizing intermediate states when studying correlation functions in NTES and TES for other models and quench protocols, such as the sinh-Gordon model which is the relativistic counterpart of the Lieb-Liniger model \cite{Kormos2009,Kormos2010}, remains open problem for future research.

\section*{Acknowledgments}
YYC and SC contributed equally.
The authors thank Chushun Tian, Shizhong Zhang, Volker Meden, Giuseppe Mussardo, and Hanns-Christoph N\"{a}ger for helpful discussions.
This work is supported by National Natural Science Foundation of China Grants No. 12574300, 12547107, and HK GRF Grants No. 17306024, CRF Grants No. C7015-24GF and C4050-23GF, CRS\_HKU701/24, and a RGC Fellowship Award No. HKU RFS2223-7S03.

\onecolumngrid

\begin{figure}[th]
	\includegraphics[width=1.1\linewidth]{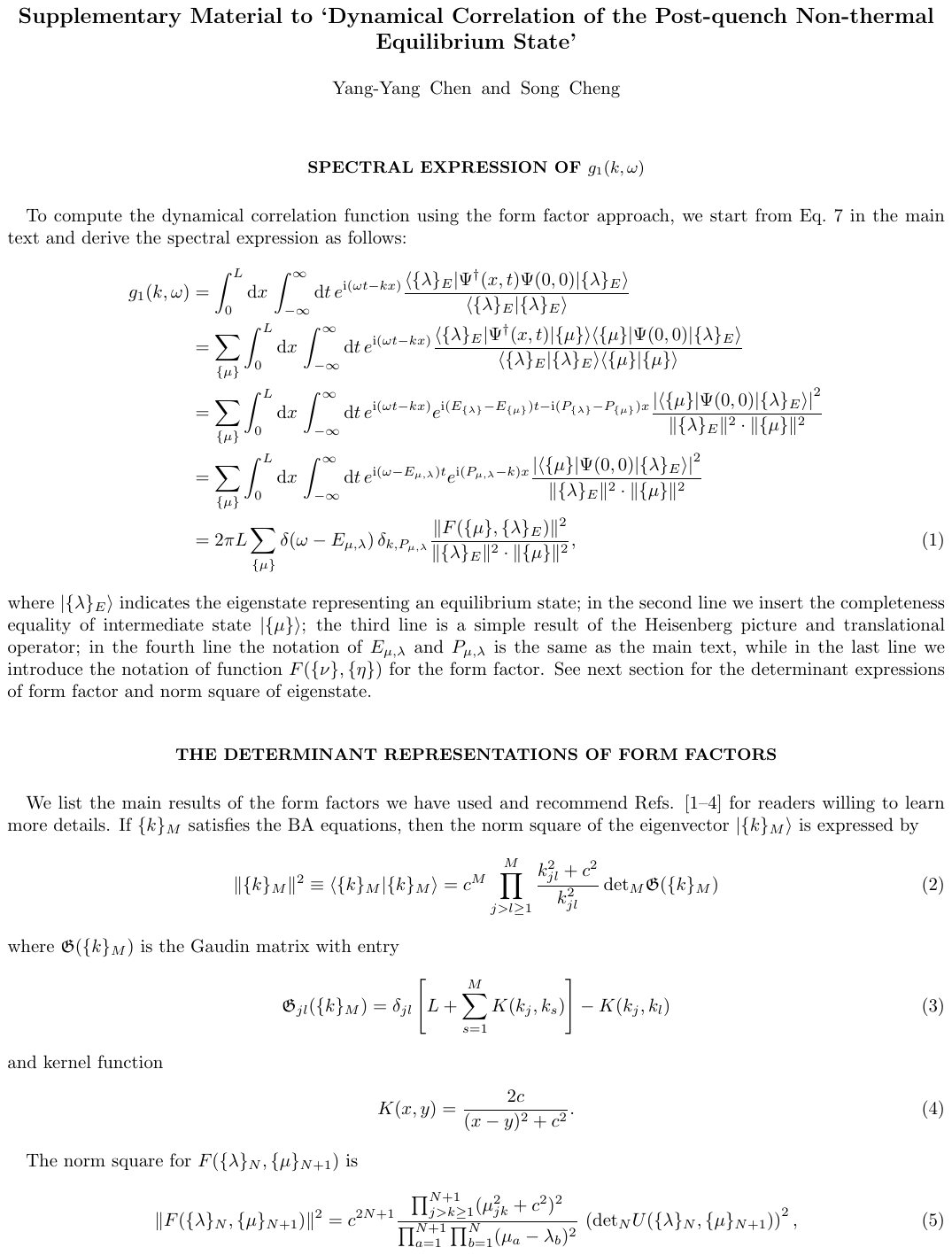}
\end{figure}

\begin{figure}[th]
	\includegraphics[width=1.1\linewidth]{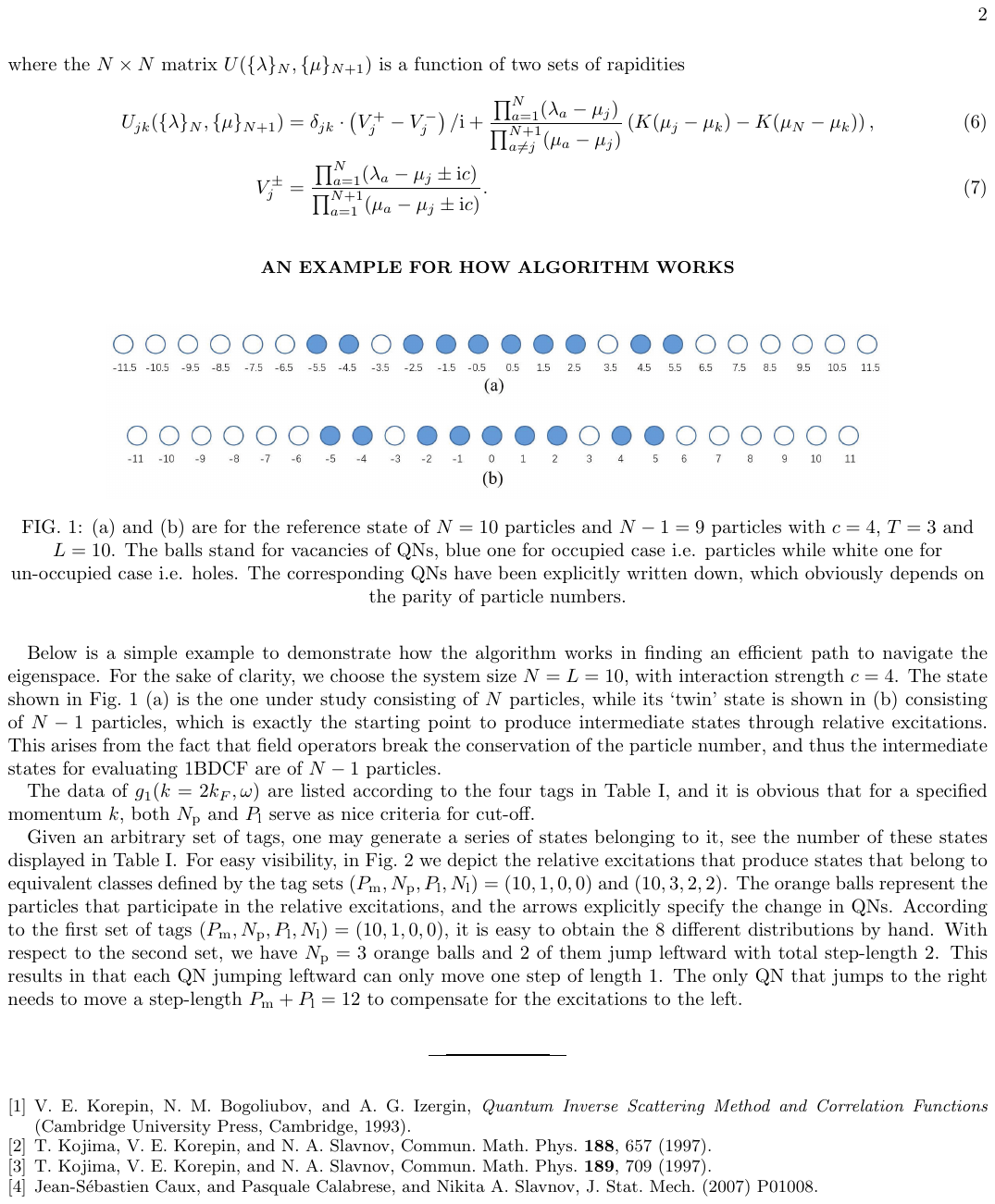}
\end{figure}

\begin{figure}[th]
	\includegraphics[width=1.1\linewidth]{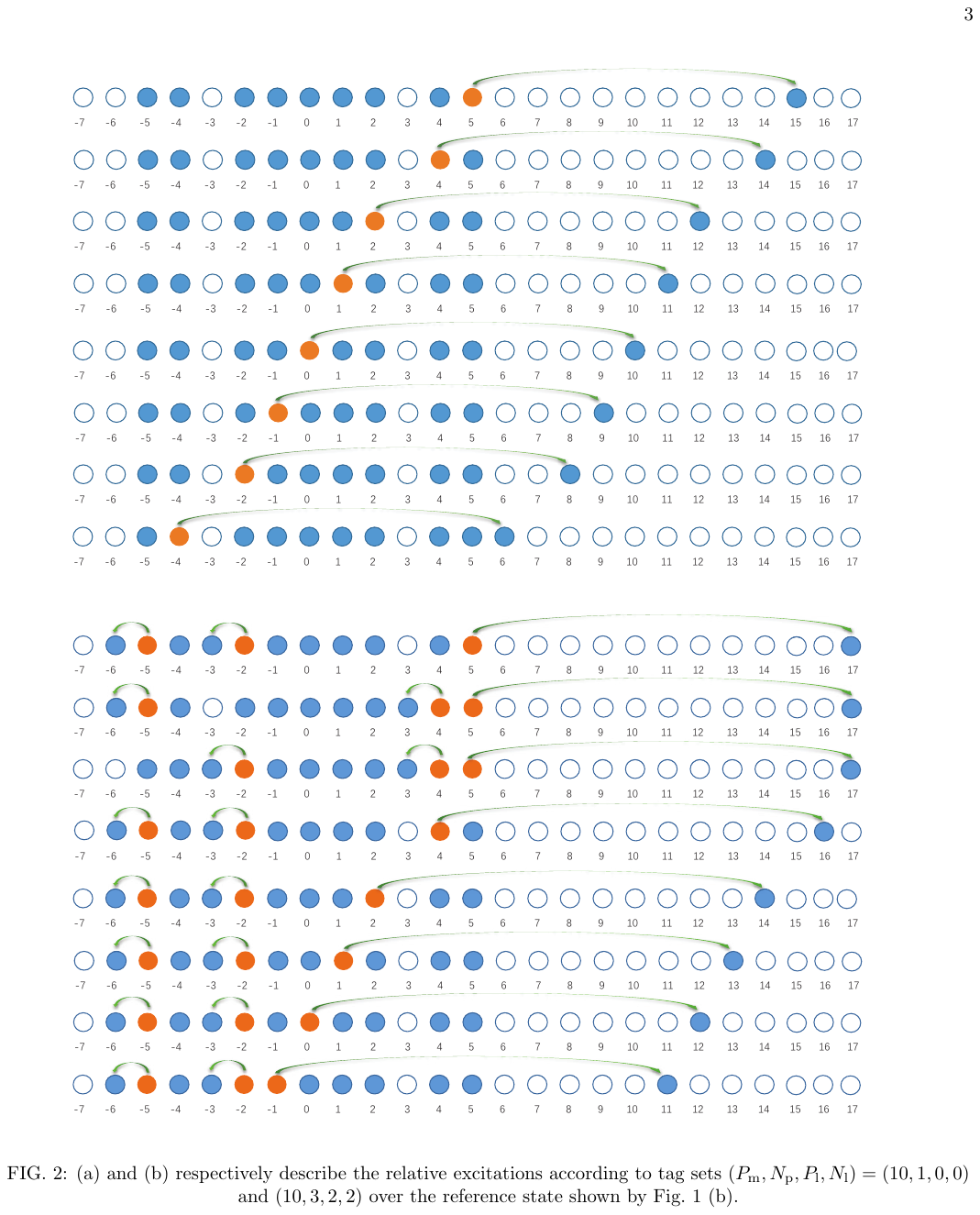}
\end{figure}

\begin{figure}[th]
	\includegraphics[width=1.1\linewidth]{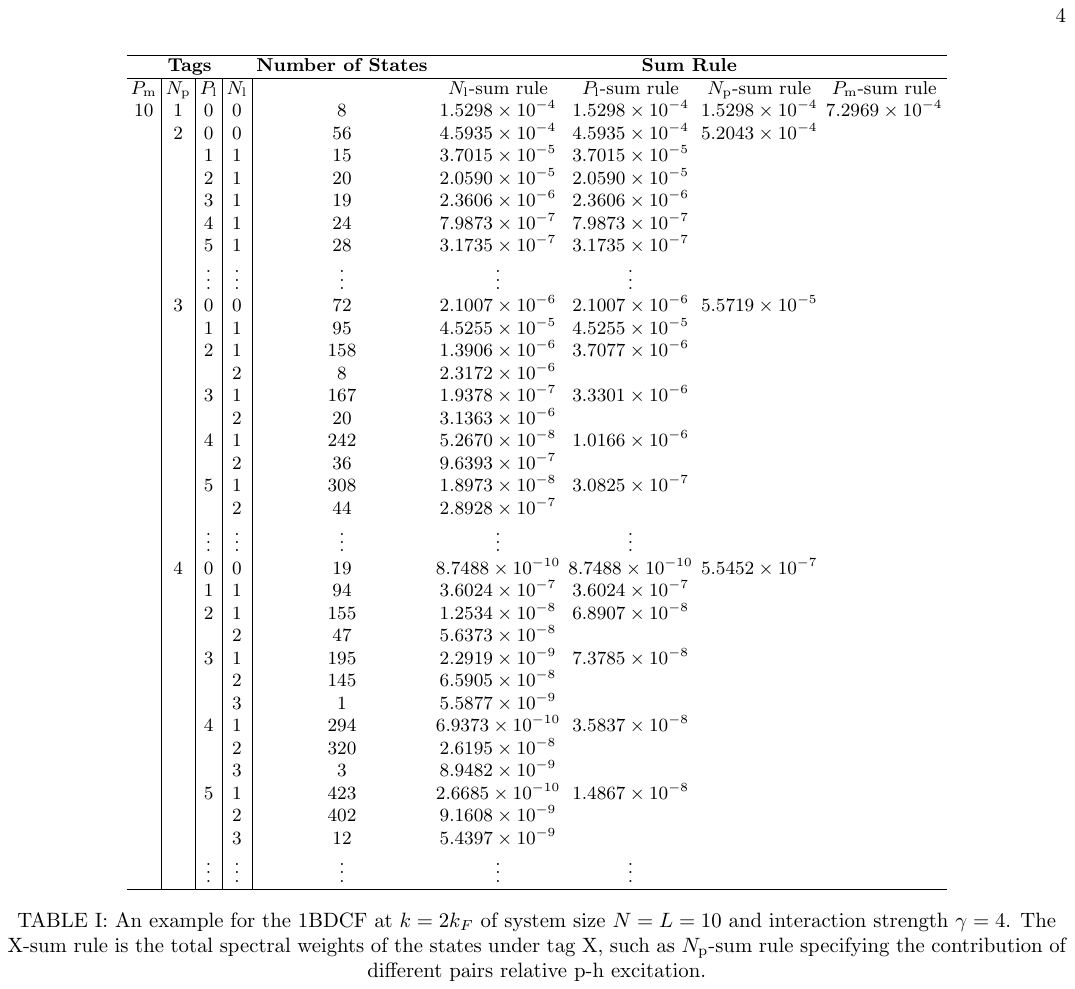}
\end{figure}

\end{document}